# Powerful Solar Flares of September 2017: Correspondence Between Parameters of Microwave Bursts and Proton Fluxes near Earth

Ilya M. Chertok

Pushkov Institute of Terrestrial Magnetism, Ionosphere and Radio Wave Propagation (IZMIRAN), Troitsk, Moscow, 108840 Russia

A strong outburst of solar activity occurred on 2017 September 4–10, when approaching a minimum of Cycle 24. It was due to the sharp development and evolution of the active region 12673 during its passage across the western half of the visible disk. Numerous strong flares occurred at this time, including 27 M-class and 4 X-class flares. Among the latter, the X9.3 flare on September 6 was the most powerful in the last 10 years. It is not surprising that this activity and associated space weather disturbances arouse great interest, and many publications have already been devoted to their study (see, e.g., Yang et al. (2017), Wang et al. (2018) and references therein).

In this note, we consider radio characteristics of three proton flares that caused discrete enhancements of solar energetic particles (SEPs) near Earth. We will proceed from the notions that, as shown, for example, in Akinyan et al. (1980a, 1980b), Chertok (1982, 1990), the flux density and frequency spectrum of microwave bursts, although the latter are generated by electrons propagating to the photosphere, reflect the number and energy spectrum of accelerated particles, including the 10–100 MeV protons coming to Earth.

The analysis is based on data of the USAF Radio Solar Telescope Network (RSTN)[1] providing round-the-clock observations of metric dynamic spectra and radio fluxes at several fixed frequencies in the range from 245 MHz to 15.4 GHz and on the *GOES*13[2] measurements of proton fluxes in three energy channels >10, 50 and 100 MeV.

In Figure 1(a), where the proton time profiles are shown, the SEPs under consideration are marked by vertical arrows. These SEPs were caused by the following flares: M5.5 flare on September 4, soft X-ray peak time 20:33 UT, coordinates S06W12; X9.3 flare on September 6, 12:02, S09W38; X8.2 flare on September 10, 16:06, S09W85. By data of the *SOHO*/LASCO coronagraph[3] just these three flares were followed by large halo coronal mass ejections (CMEs). Shown in Figure 1(b) are frequency spectra of the peak radio flux density of these proton flares.

The SEP event of September 4 was characterized by two peculiarities. First, a rather noticeable enhancement of the proton flux was observed only in the >10 MeV channel, where J10 reached ≈100 pfu (1 pfu = 1 particle cm$^{-2}$ s$^{-1}$ sr$^{-1}$). At the same time, an increase of the proton flux with $E > 50$ and 100 MeV was quite small. It means that this SEP had a very soft (steeply falling) energy spectrum with the power-law index γ ≈3.0. Second, the growth phase of the >10 MeV proton flux was prolonged and its first maximum occurred ≈10 hr after the flare. Both of these peculiarities can be to some extent due to particle propagation from this western, but near-central flare on heliolongitude W12. However, the character of the microwave radio burst indicates that the parameters of particle acceleration played the main role here. Its frequency spectrum displays a clear decrease of the radio flux in the range from 1 to 9 GHz and therefore it is soft. Flares with such a radio spectrum have mainly a post-eruption origin and are indeed accompanied by SEPs with a soft energy spectrum and prolonged growth phase.

---

[1] ftp://ftp.sec.noaa.gov/pub/warehouse/2017/2017_events/
[2] ftp://ftp.swpc.noaa.gov/pub/warehouse/2017/2017_plots/proton/
[3] https://cdaw.gsfc.nasa.gov/CME_list/halo/halo.html

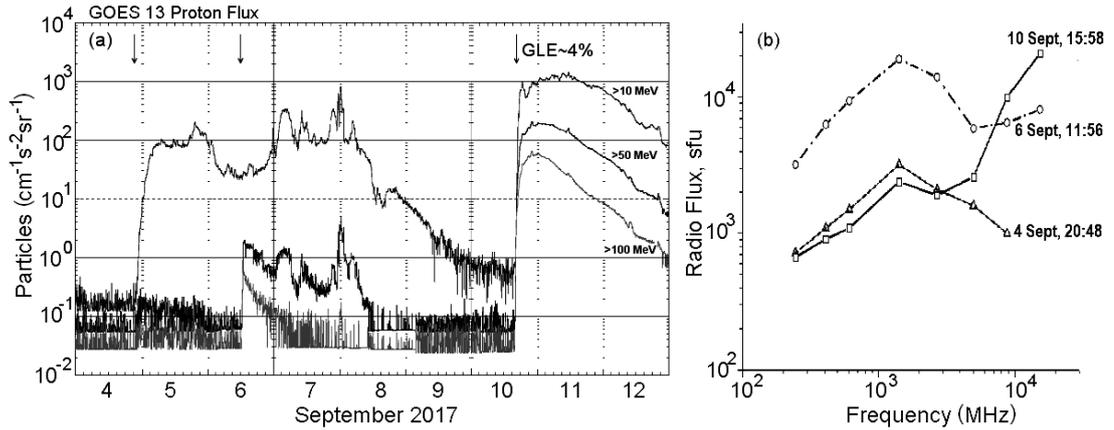

Figure 1. (a) The GOES time history of proton fluxes of 2017 September 4–12. Vertical arrows indicate three analyzed SEPs. (b) Frequency spectra of the associated radio bursts.

The September 10 event has quite different characteristics. Its radio spectrum shows a sharp increase from 3 to 15 GHz and should be considered as hard. The maximum radio flux at 15 GHz is very large $S_{15} \approx 21{,}000$ sfu (1 sfu = $10^{-22}$ W m$^{-2}$ Hz$^{-1}$). In accordance with these radio parameters, the observed SEP, as a typical western event, was very intense ($J_{10} \approx 1000$ pfu) and had a rather hard proton energy spectrum with $\gamma \approx 1.4$. Due to this, it was registered by neutron monitors as a ground level enhancement.

The September 6 event has intermediate features both on the radio spectrum and on the proton flux parameters. Its radio spectrum indicates a decimetre portion and an increasing microwave component with the peak flux at 15 GHz $S_{15} \approx 8100$ sfu. This corresponds to the observed SEP of a rather hard energy spectrum with the index $\gamma \approx 1.5$ estimated by the flux in the >50 and 100 MeV channels. Before and after this SEP, the >10 Mev proton flux was disturbed by CME from the September 4 flare and the subsequent geomagnetic storm.

The outlined results evidence once again that the intense flare microwave radio bursts contain important information about SEPs, in particular about the energy spectrum and the scale of proton fluxes coming to Earth. This can be used for diagnostics of proton flares and space weather forecasting.

A more detailed analysis of the radio bursts and SEPs of 2017 September is presented in Chertok (2018).

The author thanks the NOAA/SWPC GOES and USAF RSTN teams for data used in the study. This research was partially supported by the Russian Foundation of Basic Research under grant 17-02-00308.

## References

Akinyan S. T., Fomichev V. V. and Chertok I. M. 1980a Solar-terrestrial Prediction Proc. Washington, DC: US Department of Commerce, 1,D7

Akinyan S. T., Fomichev V. V. and Chertok I. M. 1980b Ge&Ae 20 385  ADS

Chertok I. M. 1982 Ge&Ae 22 182  ADS

Chertok I. M. 1990 AN 311 379  ADS

Chertok I. M. 2018 Ge&Ae in press

Wang H., Yurchyshyn V., Liu Ch. et al 2018 RNAAS 2 8   [ADS]

Yang S., Zhang J., Zhu J. and Song Q. 2017 ApJL 849 L21   [ADS]